\documentstyle[12pt]{article}
\begin{document}
\vspace*{3cm}
\begin{flushleft}
\large
A.N.~Storozhenko, D.S.~Kosov$^*$, A.I.~Vdovin

\vspace{2cm}
{\bf LIPKIN -- MESHKOV -- GLICK MODEL AT FINITE
TEMPERATURE }

\vspace{3.5cm}
Submitted to "Yadernaya Fizika"
\end{flushleft}

\vspace{6cm}
\noindent
\rule{8cm}{0.6pt}

\noindent
{\large $^*$ Institute for Theoretical Physics, University of Tuebingen,
Germany}

\newpage

\section{Introduction}

New approximate methods of nuclear structure theory are usually examined
by applying them to simple exactly soluble models in order to gain some
insights on a range of their validity.  One of the widely used models
is the two level schematic shell model which possess the $SU(2)$ symmetry
and is often called   $SU(2)$ or Lipkin -- Meshkov -- Glick (LMG) model [1].
Numerous applications of the LMG- model can be found in~[2].

During the last years the model has been used many times to justify approximate
methods of the many-body theory at finite temperature [3-9]. These methods
are especially interesting in view of current intensive studies of hot nuclear
systems. Previous works on the LMG- model at finite temperature [3,4]
have focused on boson expansion methods and symmetry breaking in hot nuclei.
The mixed state representation has been formulated and then applied to
the LMG- model in refs.[5-8]. In particular, the thermal Hartree - Fock approximation
as well as the thermal random phase approximation (TRPA) were studied
within the approach [8].  The thermal Hartree - Fock approximation
and the static path approximation were analyzed within
the model in ref.[9].

A new approximate method has been recently proposed [10] to describe collective
excitations in hot finite Fermi systems. This method, the so-called
renormalized TRPA (TRRPA), is an extension of the renormalized RPA of
Ken-ji Hara [11] and D.~Rowe [12] to finite temperatures.
Within TRRPA vibrational
excitations are supposed to be harmonic like in TRPA but a temperature -
dependent ground state is treated in a more consistent manner.
Namely, a finite number of thermal quasiparticles are presented in this
ground state.

In the present paper, we investigate the accuracy of the thermal
renormalized random phase approximation by comparing
it with the exact calculations for the grand canonical ensemble
for the LMG- model. Moreover, a comparison with the thermal
mean field approximation (TMFA) and TRPA is also done.

\section{The LMG- model and the grand canonical ensemble calculations}

The following version of the LMG- model is used: $N$ fermions are
distributed over two levels, each having a degeneracy $\Omega$.
The distance between the levels is $\varepsilon$, the coupling
constant $V$ does not depend on any quantum number. At  $T = 0$
and $V = 0$ the lower
level is full, the upper - empty, i.e. $N = \Omega$. The model
Hamiltonian has the form
$$
H= \varepsilon \hat{J}_{z} -\frac{1}{2} V \left(\hat{J}_{+}\hat{J}_{+}
+ \hat{J}_{-}\hat{J}_{-} \right)\; ,
\eqno(1)
$$
where the operators of quasispin  $\hat{J}$ and its projections
 $\hat{J}_{+}, \hat{J}_{-}, \hat{J}_{z}$ are defined as follows:
$$
{\hat J}^2 = \frac{1}{2} \left( \hat J_{+}{\hat J}_{+} +
{\hat J}_{-}{\hat J}_{-} \right) + \hat{J}^2_{z} \;,
$$
$$
J_{z}=\frac{1}{2} \sum_{p=1}^{\Omega}
\left( a^{+}_{2 p} a_{2 p} - a^{+}_{1 p} a_{1 p}\right) \,, \quad
\hat{J}_{+}=\sum_{p=1}^{\Omega} a^{+}_{2 p}a_{1 p} \,, \quad
\hat{J}_{-} = \left(\hat{J}_{+}\right)^{+} \,.
$$
Here $a^{+}_{i p}$ and $a_{i p}$  are particle creation and annihilation
operators on the lower  ($i=1$) or the upper ($i=2$) level.

The operators
$\hat{J}_{\pm}$ and $\hat{J}_{z}$ form  SU(2) algebra, and the quasispin operator
commutes with $H$. So the Hamiltonian matrix breaks up into submatrices $\Theta_{J}$
of dimension $2J + 1$. The Hamiltonian can be diagonalized in each of these
subspaces independently. The corresponding eigenvalues are denoted by
$E^{J}_{1}, E^{J}_{2},...E^{J}_{2J+1}$.

To calculate the grand canonical partition function, one needs the eigenvalues
$E^{J}_{k}$ and the degeneracies of irreducible quasispin representations
$\Theta_{J}$ for different particle numbers from the range $0 < N \le 2\Omega$.
The latter have been determined in ref.[7], and here we use
this result. The whole number of the ensemble states, i.e., the whole number
of the eigenstates of the LMG- systems formed by two $\Omega$- degenerated
levels with a number of particles varying from $1$ to $2\Omega$ is equal to
$2^{2\Omega}$. Any state of the ensemble can be written in the following
form:
$$
|g_{1}p_{1}, g_{2}p_{2},...g_{n}p_{n}\rangle =
a^{+}_{g_{1}p_{1}} a^{+}_{g_{2}p_{2}},...a^{+}_{g_{n}p_{n}} |0\rangle \,,
\qquad  a_{gp}|0\rangle = 0 \, ,
$$
The indices $g_{i}, p_{i}, n$ have the following meanings:
$$
g_{i} \in \{1,2\},\; p_{i} \in \{1,...\Omega\}, \;
i \in \{1,...n\}, \; n \in \{1,...2\Omega\},
$$
i.e., $g$ marks the lower and the upper levels, $p$ -- sublevels,
$i$ is an index of a particle and $n$ is the particle number in the
particular LMG- system from the grand canonical ensemble. If $n = 0$,
$|g_{1}p_{1}, g_{2}p_{2},...g_{n}p_{n}\rangle = |0\rangle $.
A particular distribution of the given number of particles over two
degenerate levels can be characterized by numbers $\nu_1$ and $\nu_2$:
$\nu_1$ is a number of sublevels  which
are occupied by particles for both the lower and upper levels;
$\nu_2$ is a number of sublevels which are occupied for neither the lower
nor the upper level. The quasispin $J$ of the state is determined by the
distribution of the rest of particles over $2\tau$ sublevels where
$2\tau = \Omega - \nu_1 - \nu_2$. The number $2(\tau + \nu_1)$ is equal to
the number of particles. We denote by
$\Gamma_{p_{1}, p_{2},...p_{2\tau + \nu_{1}}}$ the subspace of states with
$\nu_1$ occupied and $\nu_2$ empty sublevels. Its dimension is $2^{2\tau}$.
There exist $\Omega!/(2\tau)! \nu_{1}! \nu_{2}!$ distinct subspaces
$\Gamma_{p_{1}, p_{2},...p_{2\tau + \nu_{1}}}$ for fixed $\tau$ and $\nu_1$.
Each of them may be decomposed into irreducible subspaces with fixed
quasispin values $\Theta_{\tau}$ (appearing once), $\Theta_{\tau - 1}$
(appearing $g^{\tau}_{1}$ times),
$\Theta_{\tau -2}$ (appearing $g^{\tau}_{2}$ times), ... ,
$\Theta_{\tau - k}$ (appearing $g^{\tau}_{k}$ times),... ,
$\Theta_{\tau - [\tau]}$ (appearing $g^{\tau}_{[\tau]}$ times). Here
$$
g^{\tau}_{k} = \frac{(2\tau)!}{k! (2\tau - k)!} -
\frac{(2\tau)!}{(k - 1)! (2\tau - k + 1)!} .
$$
and
$[\tau] = \tau$, if $\tau$ is integer, $[\tau] = \tau - 1/2$ if $\tau$
is half-integer.

Then, the exact grand partition function is
$$
Z(T) = \sum_{\tau \nu_{1} \nu_{2}} \frac{\Omega!}{(2\tau)! \nu_{1}! \nu_{2}!}
\sum_{k} g^{\tau}_{k}
\sum_{m} \exp\left[ -
\frac{E^{\tau - k}_{m} - 2(\tau + \nu_{1})\lambda}{T}\right]
$$
The expressions for average energy, quasispin z-projection and the total
fermion number are
\begin{eqnarray}
\langle H \rangle_{GCE}& = &
\frac{1}{Z} \sum_{\tau \nu_{1} \nu_{2}}
\frac{\Omega!}{(2\tau)! \nu_{1}! \nu_{2}!}
\sum_{k} g^{\tau}_{k}
\sum_{m} E^{\tau - k}_{m} \exp\left[ -
\frac{E^{\tau - k}_{m} - 2(\tau + \nu_{1})\lambda}{T}\right] \nonumber \\
&&\nonumber\\
\langle \hat{J}_{z} \rangle_{GCE}& = &
\frac{1}{Z} \sum_{\tau \nu_{1} \nu_{2}}
\frac{\Omega!}{(2\tau)! \nu_{1}! \nu_{2}!}
\sum_{k} g^{\tau}_{k}
\sum_{m} \langle k, \tau |\hat{J}_{z}|k, \tau \rangle  \exp\left[ -
\frac{E^{\tau - k}_{m} - 2(\tau + \nu_{1})\lambda}{T}\right] \nonumber \\
&&\nonumber\\
\langle \hat N \rangle_{GCE} & = &
\frac{1}{Z} \sum_{\tau \nu_{1} \nu_{2}}
\frac{\Omega!}{(2\tau)! \nu_{1}! \nu_{2}!} 2\left( \tau + \nu_{1}\right)
\sum_{k} g^{\tau}_{k}
\sum_{m} \exp\left[ -
\frac{E^{\tau - k}_{m} - 2(\tau + \nu_{1})\lambda}{T}\right]
\nonumber
\end{eqnarray}

\section{Thermo field dynamics: basic elements}

To be more understandable while describing approximate methods,
we briefly recapitulate the formalism of thermo field dynamics (TFD)
(see, [3, 13-15]).

The extension of quantum field theory to finite
temperature requires the field degrees of freedom to be doubled.
In TFD, this doubling  is achieved by introducing an additional tilde space.
A tilde conjugate operator $\tilde A$ is assigned to an operator
$A$ (acting in ordinary space) through the tilde conjugation rules
 $$ \widetilde{(AB)} = \tilde A \tilde B \,; \quad
 \widetilde{(a A + b B)} = a^* \tilde A + b^* \tilde B \; ,
$$
where $A$ and $B$ represent ordinary operators and $a$ and $b$ are c-numbers.
The asterisk denotes the complex conjugation. The tilde operation commutes
with hermitian conjugation and any tilde and non-tilde
operators are assumed to commute or anticommute with each other.
A double application of tilde operation changes a sign of a fermionic
operator and saves it for a bosonic one. The whole Hilbert space of a
heated system is a direct product of ordinary and tilde spaces.
A formal quantity playing a central role in the present discussion
is the so-called thermal Hamiltonian:
$$
{\cal H} = H - \tilde H
$$
The operator ${\cal H}$ serves to translate
temperature dependent wave functions along the time axis.
It means that an "excitation spectrum" of a hot system
(or, in other words, a set of energies corresponding to the thermal
equilibrium states) should be obtained by the diagonalization of $\cal H$.

The temperature-dependent vacuum $|\Psi_{0} (T) \rangle $ is the
eigenvector of ${\cal H}$ with eigenvalue~$0$
$$ {\cal H}|\Psi_{0}(T)\rangle =0 $$
If one determines the thermal vacuum state as
$$|\Psi_{0} (T) \rangle = \frac{1}{\sqrt{Tr(\exp(- H/T ))}} \sum_{n}
\exp(-\frac{E_{n}}{2T})|n\rangle \otimes |\tilde{n}\rangle $$
where $E_{n}, |n\rangle$ and $|\tilde n\rangle$ are eigenvalues,
eigenvectors and their tilde counterparts of the Hamiltonian $H$,
respectively,
the expectation value $\langle \Psi_{0} (T)|O|\Psi_{0} (T) \rangle$
will exactly correspond to the grand canonical ensemble average
$\ll O \gg $ of a given observable~$O$.

In practice, it is impossible to find the exact thermal vacuum for
a full Hamiltonian of a many-body system. In setting up approximate
schemes, the usual starting point is the thermal mean-field approximation.
In this case, the thermal vacuum $|\Psi_{0} (T) \rangle $ is
an eigenvector of the uncorrelated thermal Hamiltonian
$$
{\cal H}_{MF} |0(T)\rangle =
(H_{MF} - \tilde{H}_{MF})|0(T)\rangle = \sum_{i}\varepsilon_{i}(a^{+}_{i}a_{i}
-\tilde{a}^{+}_{i} \tilde{a}_{i}) |0(T)\rangle   = 0 \, . \eqno(2)
$$
The solutions  of eq.~(2) define the vacuum $|0(T)\rangle$ for so-called
thermal quasiparticles $\beta,\tilde{\beta}$:
$$ \beta_{i} =
x_{i}a_{i} - y_{i}\tilde{a}^{+}_{i} $$
$$\tilde{\beta}_{i} =
x_{i}\tilde{a}_{i} + y_{i} a^{+}_{i}
$$
$$ \beta_{i} |0 (T) \rangle  =
\tilde{\beta}_{i} |0 (T) \rangle  = 0 \, ,
$$
where the coefficients $x_i , y_i$ denote the thermal Fermi
occupation probabilities of the states $a^{+}_{i} |0\rangle$
($|0\rangle$ is a vacuum for $a_{i}$)
$$ x_{i} = \sqrt{1-f_{i} }\; , \; y_{i}= \sqrt{f_{i} }  $$
$$ f_{i} = \frac{1}{1+\exp(\varepsilon_{i}/T)}$$
Sometimes the $\{x, y\}$ transformation is called the thermal
Bogoliubov transformation. It is a unitary transformation and thus
conserves the commutation relations.

\section{Approximate methods}

Now we apply the TFD formalism to the LMG- model and derive
the corresponding equations of TRRPA. A more general formulation of
the thermal renormalized random phase approximation can be found in
refs.[10,16,17]. Moreover, within the Hartree --
Fock method, depending on the value of the coupling constant $V$
two different phases of the LMG- system exist -- a normal phase and a
deformed one. The present consideration is restricted to a normal phase.
So we do not take into account the mean field rearrangement which occurs
if the value of the effective coupling constant $\chi = V(N - 1)/\varepsilon$
becomes more than unity.

The model thermal Hamiltonian
${\cal H} = H - \tilde{H}$, where $H$ has the form (1), has to be written
in terms of the thermal quasiparticle operators. The first item in (1)
conserves the diagonal form. The interaction operator becomes more complicated.
For further studies we need only that part of $\cal H$ which consists of
the terms with even numbers of both creation and annihilation thermal
quasiparticle operators. Namely,
$$
{\cal H}' = \varepsilon \left( B - \tilde B \right)
- \frac{V(f_{1} - f_{2})}{2}
\left[ \left( A^{+}{}^{2} + A^{2} \right) -
\left({\tilde A}^{+}{}^{2} + {\tilde A}^{2} \right)\right] \,,
\eqno(3)
$$
where
$$
B = \frac{1}{2} \sum_{p=1}^{\Omega}\left(\beta^{+}_{2p} \beta_{2p}
- \beta^{+}_{1p} \beta_{1p}\right) \,, \quad
A^{+} = \sum_{p=1}^{\Omega} \beta^{+}_{2p} \tilde{\beta}^{+}_{1p}
 \,,\quad
{\tilde A}^{+} = \sum_{p=1}^{\Omega} \beta^{+}_{1p} \tilde{\beta}^{+}_{2p}
\,.
$$

The following exact commutation rules are valid for the thermal
biquasiparticle operators $A$, $A^{+}$, $\tilde A$ and ${\tilde A}^{+}$:
$$
\left[ A, A^{+} \right] = N -
\sum_{p=1}^{\Omega} \tilde{\beta}_{1p}^{+}\tilde{\beta}_{1p} -
\sum_{p=1}^{\Omega} \beta_{2p}^{+}\beta_{2p}\,, \quad
\left[ \tilde{A}, {\tilde A}^{+}\right] = N -
\sum_{p=1}^{\Omega} \beta_{1p}^{+} \beta_{1p} -
\sum_{p=1}^{\Omega} \tilde{\beta}_{2p}^{+} \tilde{\beta}_{2p}\,.
\eqno(4)
$$
All other commutators between the operators $A$, $A^{+}$, $\tilde A$ and
${\tilde A}^{+}$ vanish.

By the use of the Wick theorem one can approximate [10,16]
the r.h.s. of (4) by c-numbers
neglecting an influence of the pairs of normal ordered
operators $: \beta^{+} \beta :$ and
$:\tilde{\beta}^{+} \tilde{\beta}:$. Namely,
$$
\left[ A, A^{+}\right] =
\left[ \tilde A, {\tilde A}^{+}\right] =
N \left(1-\rho_1 -\rho_2\right)
\equiv N \left( 1- 2\rho \right) \,.
\eqno(5)
$$
Here $\rho_{i}$ are the numbers of thermal quasiparticles in
the temperature - dependent ground state $|\Psi_{0}(T)\rangle$ that will
be defined later on. That is
$$
\rho_{i} = \frac{1}{N}\langle \Psi_{0}(T)|N_{i}^{\beta} |\Psi_{0}(T)\rangle =
 \frac{1}{N}\langle \Psi_{0}(T)|\tilde{N}_{i}^{\beta} |\Psi_{0}(T)\rangle \,
$$
where $N_{i}^{\beta}$ is the operator of the number of thermal quasiparticles
$ N_{i}^{\beta} = \sum_{p=1}^{\Omega} \beta_{ip}^{+}\beta_{ip}\,.$

The thermal Hamiltonian (3) can be diagonalized in the space of two one-phonon
states constructed as bilinear forms of the thermal biquasiparticle operators:
$$
Q^{+}_{1} |\Psi_{0}(T)\rangle
= \left( \psi_{1} A^{+} - \phi_{1} A \right)|\Psi_{0}(T)\rangle
\eqno(6) $$
$$
Q^{+}_{2} |\Psi_{0}(T)\rangle
= \left( \psi_{2} {\tilde A}^{+} - \phi_{2} \tilde A \right)|\Psi_{0}(T)\rangle \,.
$$
Now we define the wave function of the temperature - dependent ground state
$|\Psi_{0}(T)\rangle$ as the thermal phonon vacuum,
i.e. $Q_{1,2} |\Psi_{0}(T)\rangle = 0$.

The states (6) have to be orthonormal. Thus, taking account of eq.~(5)
the following constraints on the amplitudes $\psi$ and $\phi$ are derived
$$
 \psi^2_i - \phi^2_i = \left[ N( 1- 2\rho )\right]^{-1} \,, \quad i = 1,2 \,.
$$
The system of equations for $\psi_i$, $\phi_i$ and the phonon
frequencies $\omega_i$ is easily obtained by the equation of motion method.
It appears that only a positive value of $\omega_1$ and a negative value of
$\omega_2$ is allowed under a requirement that the wave functions
$Q^{+}_{1} |\Psi_{0}(T)\rangle$ and $ Q^{+}_{2} |\Psi_{0}(T)\rangle $ are
vectors of the Hilbert space. The eigenvalue - eigenvector problem has the
following solution:
$$
\omega_1 = \omega \equiv \sqrt{\varepsilon^2 - V^2 \left( f_2 - f_1 \right)^2
\left( 1 - 2\rho \right)^2 \left( N - 1 \right)^2 } \,,
$$
$$
\psi^2_{1} = \frac{\varepsilon + \omega}{2N\omega(1 - 2\rho)}\,, \quad
\phi^2_{1} = \frac{\varepsilon - \omega}{2N\omega(1 - 2\rho)} \,,
$$
$$
\omega_2 = - \omega \,, \quad
\psi^2_{2} = \psi^2_{1} \,, \quad \phi^2_{2} = \phi^2_{1}\,.
$$

One more equation has to be added to the above system -- the
equation for $\rho$. To evaluate this equation we need an expression for the
thermal phonon vacuum state. The latter can be derived from the thermal
quasiparticle vacuum state $|0(T)\rangle$ by  a unitary transformation
$$
|\Psi_{0}(T)\rangle = \sqrt{R} e^{S}|0(T)\rangle \,, \quad
S = \frac{1}{2(1-2\rho)}\frac{\phi_1}{\psi_1}
\left( A^{+} A^{+} + {\tilde A}^{+} {\tilde A}^{+}\right) \,.
$$
By the use of standard techniques of the operator calculus [11]
we get
$$
\rho = \frac{1}{2} \frac{\varepsilon - \omega}{ N \omega}
\eqno(7)
$$
It is interesting to note that in the thermodynamic limit, i.e. at
$N \to \infty$, $\rho$ vanishes and the TRRPA equations are reduced to the
TRPA ones.

Let us display
the expressions for the average energy, the average quasispin z-projection
and the variance of the particle number
\begin{eqnarray}
\langle \hat{H}' \rangle_{TRRPA}& = &
\frac{N\varepsilon(f_2 - f_1)}{2} \left( 1 - 2\rho\right)  +
\frac{\varepsilon^2 - \omega^2}{2\omega} \times
\frac{(f_2 - f_1)^2 + 1}{2(f_2 - f_1)}
\nonumber \\
&&\nonumber \\
\langle \hat{J}_{z} \rangle_{TRRPA}& = & \frac{N(f_2 - f_1)}{2}
\left( 1 - 2\rho\right)
\nonumber   \\
&&\nonumber \\
\Delta N_{TRRPA}& =& \sqrt{N (1 - 2\rho) \left[ f_1 (1 - f_1) +
f_2 (1 - f_2)\right]} \, .
\nonumber
\end{eqnarray}
The above expectation values were taken over the TRRPA vacuum state.

It seems appropriate to give expressions for the same quantities
within other approximations -- TRPA and TMFA.
The TRPA expressions are obtained from the TRRPA ones by putting $\rho$ = 0.
In this case, the commutator $\left[A, A^{+}\right]$ is
equal to $N$ and the expectation values are taken over the TRPA vacuum
\begin{eqnarray}
\langle \hat{H}' \rangle_{TRPA}& =&
\frac{N\varepsilon(f_2 - f_1)}{2}  +
\frac{\varepsilon^2 - \omega^2}{2\omega} \times
\frac{(f_2 - f_1)^2 + 1}{2(f_2 - f_1)} \; ,
\nonumber\\
&&\nonumber \\
\langle \hat{J}_{z} \rangle_{TRPA}& = & \frac{N(f_2 - f_1)}{2} \; ,
\nonumber\\
&&\nonumber \\
\Delta N_{TRPA}& = &\sqrt{ N \left[ f_1 (1 - f_1) + f_2 (1 - f_2)\right]} \;.
\nonumber
\end{eqnarray}

Within TMFA the interaction between particles is omitted and one has to evaluate
the expectation values over the thermal quasiparticle vacuum $|0(T)\rangle$.
The TMFA expressions for
$\langle \hat{J}_{z} \rangle_{TMFA}$ and $\Delta N_{TMFA}$ are the same as in
TRPA. For $\langle \hat{H} \rangle_{TMFA}$ one gets
$$
\langle \hat{H} \rangle_{TMFA} =
\frac{N\varepsilon(f_2 - f_1)}{2} \;.
$$

\section{Results and discussion}

The numerical calculations are done for the LMG- system with
$N = \Omega =$10 particles and $\varepsilon$=2.
The results are displayed in Figs.~1-6.

Firstly, we discuss a dependence of some characteristics of the system
on the effective coupling constant $\chi$. The energy of the collective
state $\omega$ as a function of $\chi$ at $T =$ 0 and 0.25$\varepsilon$
is displayed in Fig.~1. Besides the results of the TRPA and
TRRPA calculations the exact solution at $T =$0 is also shown. As it should be,
with increasing $\chi$ the energy $\omega$ goes down.
Within TRPA $\omega$ vanishes
at $\chi =$1. This collapse does not take place for the exact solution
as well as for the TRRPA result. This last feature of the RRPA solution
is well known in the case of a cold nucleus and is actively used in some
recent nuclear structure calculations [18]. As it has been demonstrated
for the first time in ref.[10], the same is valid at $T \neq$ 0.
In the present version of the LMG- model heating effectively  weakens
the interaction of particles (the effective coupling constant $\chi$
is multiplied by a factor of $f_1 - f_2 < 1$) and the TRPA collapse occurs at
larger $\chi$- values. In TRPA when $\chi \to \chi_{collapse}$
$\langle \hat H \rangle_{TRPA} \sim - \omega^{-1} \to -\infty $. It is not
the case for TRRPA (see Fig.~2). The value $\langle \hat H \rangle_{TRRPA}$
goes down much slower and remains even greater than the exact value
$\langle \hat H \rangle_{GCE}$. At large values of $\chi$ the strong
difference between $\langle \hat H \rangle_{TRRPA}$ and
$\langle \hat H \rangle_{GCE}$ is due to neglecting the mean field
rearranging.

In Figs.~3-5, the average energy of the system, the average
$\hat J_z$ value and the variance of the particle number as functions of $T$
are displayed ($\chi = $0,95). The noticeable difference between the exact
and the approximate values is only at moderate $T \le 0,3\varepsilon$.
Here TRRPA works evidently better than TRPA and TMFA. The absolute
values $\langle H \rangle_{TRPA}$ and  $\langle \hat J_z \rangle_{TRPA}$ are
greater than $\langle H \rangle_{GCE}$ and $\langle \hat J_z \rangle_{GCE}$,
respectively. At the same time,
$|\langle H \rangle_{TRRPA}| < |\langle H \rangle_{GCE}| $. The relation
 $|\langle \hat J_z \rangle_{TRRPA}| < |\langle \hat J_z \rangle_{GCE}| $
is valid only at $T < 0,8\varepsilon$ but then
$|\langle \hat J_z \rangle_{TRRPA}|$ appears to be slightly greater
than $|\langle \hat J_z \rangle_{GCE}|$. At $T > 0,5\varepsilon$ the
differences between the exact and the approximate results is
negligible. The difference between exact and approximate values
of the particle number variance is only 2-3\%, i.e. even less than for
other variables. Decreasing in the difference with raising up $T$
is a result of effective weakening of the interaction.

The value $\Delta N / N$ as a function of $N$ is shown in Fig.~6.
It decreases slowly when $T$ increases, and its typical value at
$N =$ 10-30 is around 10\%. The approximate methods disturb
$\Delta N$ only slightly. The difference between different approximations
is of minor importance although formally TRRPA seems to be better.

\section{Summary}

Taking the Lipkin -- Meshkov -- Glick model as an example we have
studied a validity of some approximate methods of many-body theory at
finite temperature.
The average energy, the average quasispin z-projection and the particle
number variance as functions of temperature and particle number have
been calculated in different approximations as well as exactly
with the grand canonical partition function. On the whole, TRRPA gives
better results than other approximations. Its advantages are especially
evident at moderate temperatures $T \le 0,5\varepsilon$. With increasing
$T$ and $N$, results of approximate methods improve rapidly and
at $T \gg \varepsilon$ the difference between exact and
approximate results is invisible.

In the present paper, we have studied only the case with not too strong
particle interaction ($\chi < 1$). Investigations of the
deformed phase of the LMG- model are in progress.

The stimulating discussions with Prof.~V.V.~Voronov are acknowledged.
The work is done under the partial support of RFBR
(grant of RFBR 96-15-96729).

The investigation has been performed at the Bogoliubov Laboratory of
Theoretical Physics,  Joint Institute for Nuclear Research and
at the Institute for Theoretical Physics of the University of Tuebingen.

\newpage
\begin{flushleft}
{\large \bf References :}

~1. Lipkin H.J., Meshkov~N. and Glick~A.J.,
Nucl.Phys. 62 (1965) 188.

~2. Ring~P., Schuck~P., ``The Nuclear Many-Body Problem'',

~~~~~New York: Springer-Verlag, 1980.

~3. Hatsuda T., Nucl.Phys. A492 (1989) 187.

~4. Walet N.R., Klein~A., Nucl.Phys. A510 (1990) 261.

~5. Kuriyama~A. et al., Prog. Theor. Phys. 87 (1992) 911.

~6. Kuriyama~A. et al., Prog. Theor. Phys. 94 (1995) 1039.

~7. Kuriyama~A. et al., Prog. Theor. Phys. 95 (1996) 339.

~8. Kuriyama~A. et al., Prog. Theor. Phys. 96 (1996) 125.

~9. Tsay Tzeng S.Y., et al., Nucl.Phys. A590 (1994) 277.

10. Avdeenkov A.V., Kosov D.S., Vdovin~A.I.,
Mod. Phys. Lett. A11 (1996) 853.

11. Ken-ji Hara, Prog. Theor. Phys. 32 (1964) 88.

12. Rowe~D.J., Phys. Rev. 175 (1968) 1283.

13. Umezawa H., Matsumoto H., Tachiki M.,
 ``Thermo field dynamics and condensed

~~~~~states'', North-Holland, Amsterdam, 1982.

14. Tanabe K., Phys.Rev. C37 (1988) 2802.

15. Kosov D.S., Vdovin A.I., Phys. At. Nucl. 58 (1995) 766.

16. Vdovin A.I., Kosov D.S., Nawrocka W., Teor. Mat. Fiz.
111 (1997) 279.

17. Kosov D.S., Vdovin A.I., Wambach J.,
 e-Print LANL: nucl-th/970002, 1997.

18. Karadjov D., Voronov V.V., Catara F., Phys. Lett. 306B (1993) 197;

~~~~~Toivanen~J., Suhonen~J., Phys. Rev. Lett. 75 (1995) 410.
\end{flushleft}

\newpage
\centerline{\large Figure captions}

\begin{itemize}
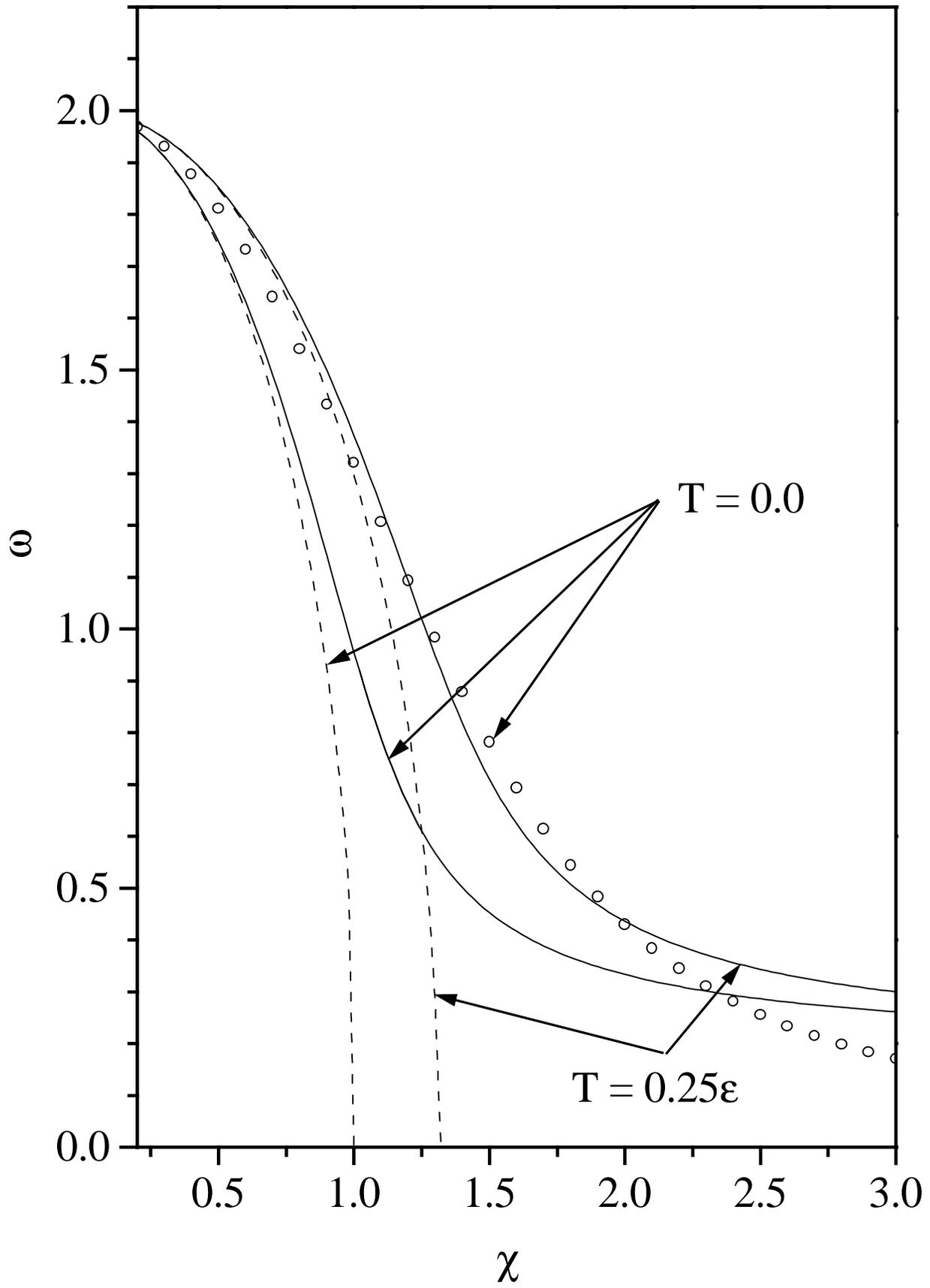
\item[Fig.~1]
The energy of the lowest excited state in the LMG- model
as a function of the effective coupling constant $\chi$
at $T =$ 0 and 0,25$\varepsilon$. The exact results
-- open circles; the TRPA (RPA) results -- dashed lines;
the TRRPA (RRPA) results -- solid lines.

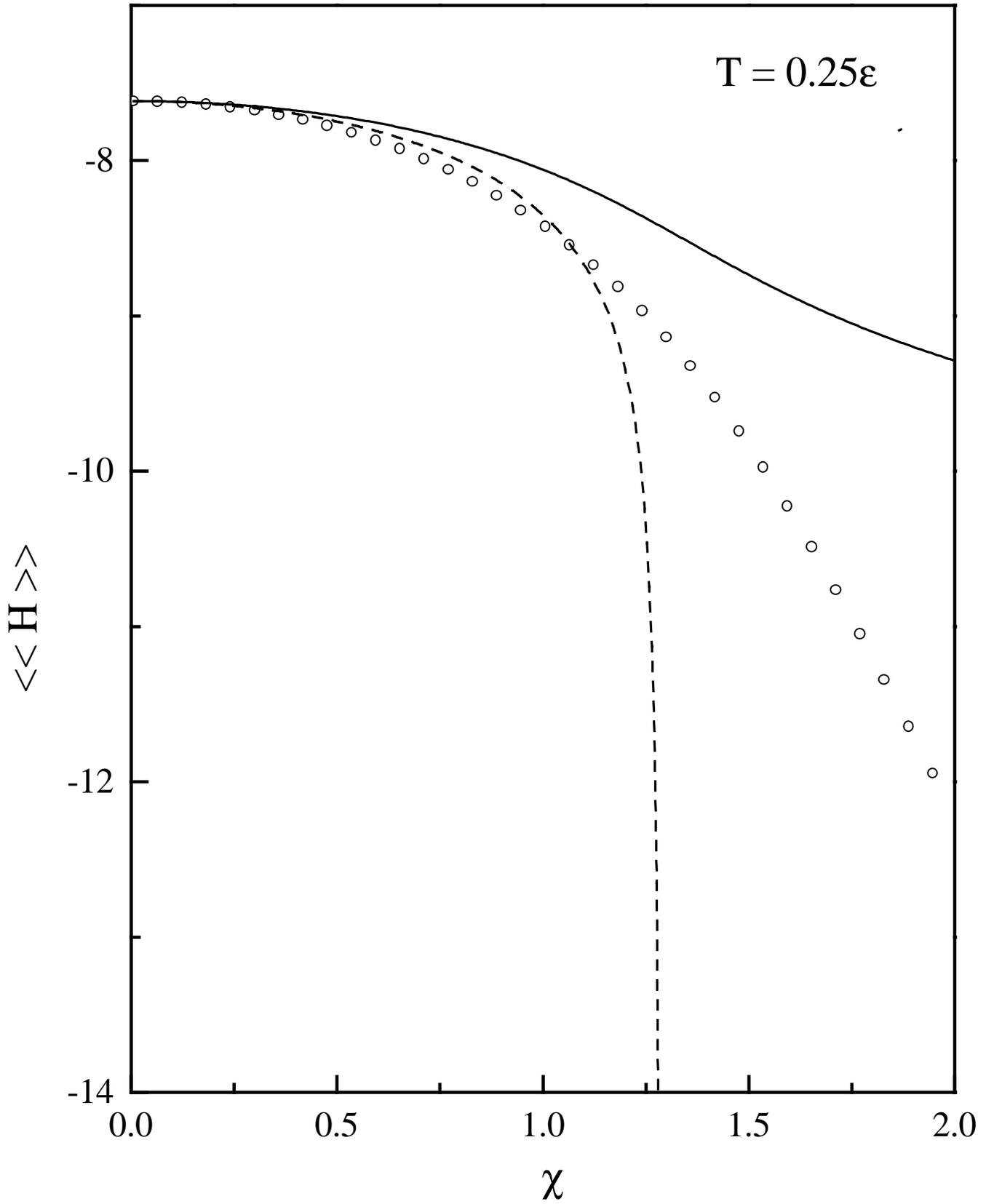
\item[Fig.~2]
The average energy of the LMG- system $\langle H \rangle$ as a function
of the effective coupling constant $\chi$.
The exact results for the grand
canonical ensemble -- open circles; the TRPA results -- dashed line;
the TRRPA results -- solid line.

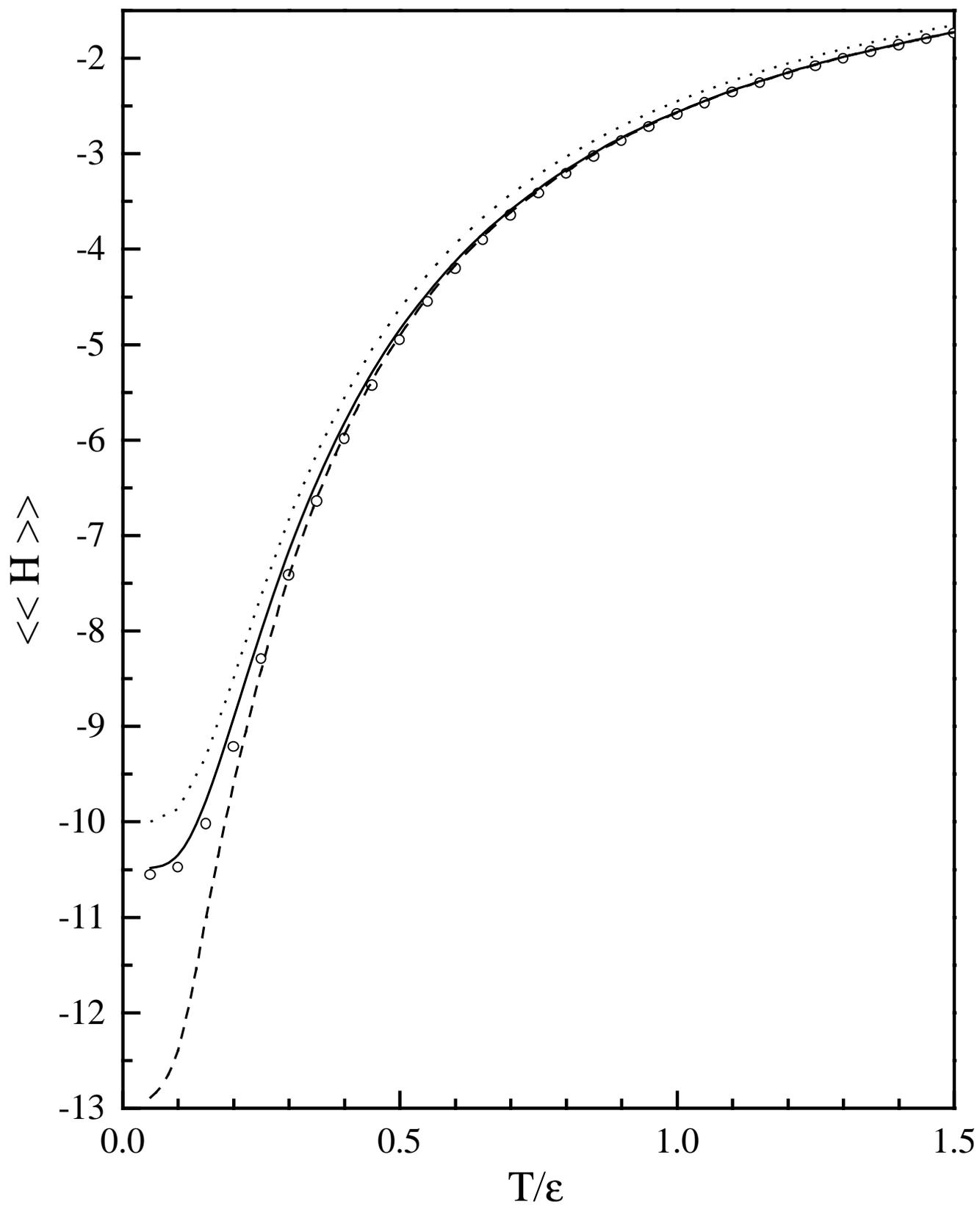
\item[Fig.~3]
The average energy $\langle H \rangle$ as a function of temperature $T$.
The exact results for the grand
canonical ensemble -- open circles; the TRPA results -- dashed line;
the TRRPA results -- solid line.

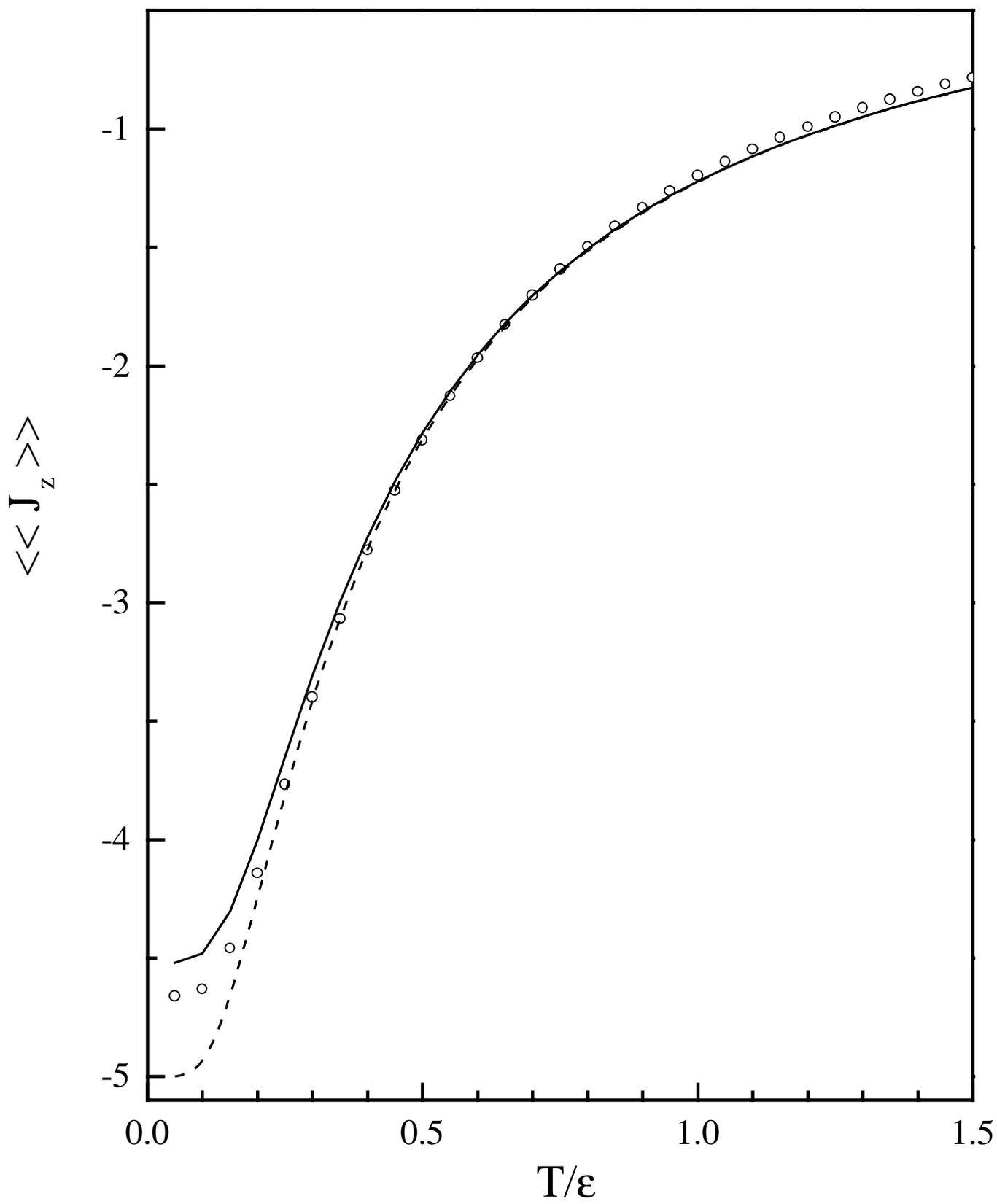
\item[Fig.~4]
The average value of the quasispin projection  $\langle \hat J_z \rangle$
as a function of temperature $T$. For notation see Fig.~3.

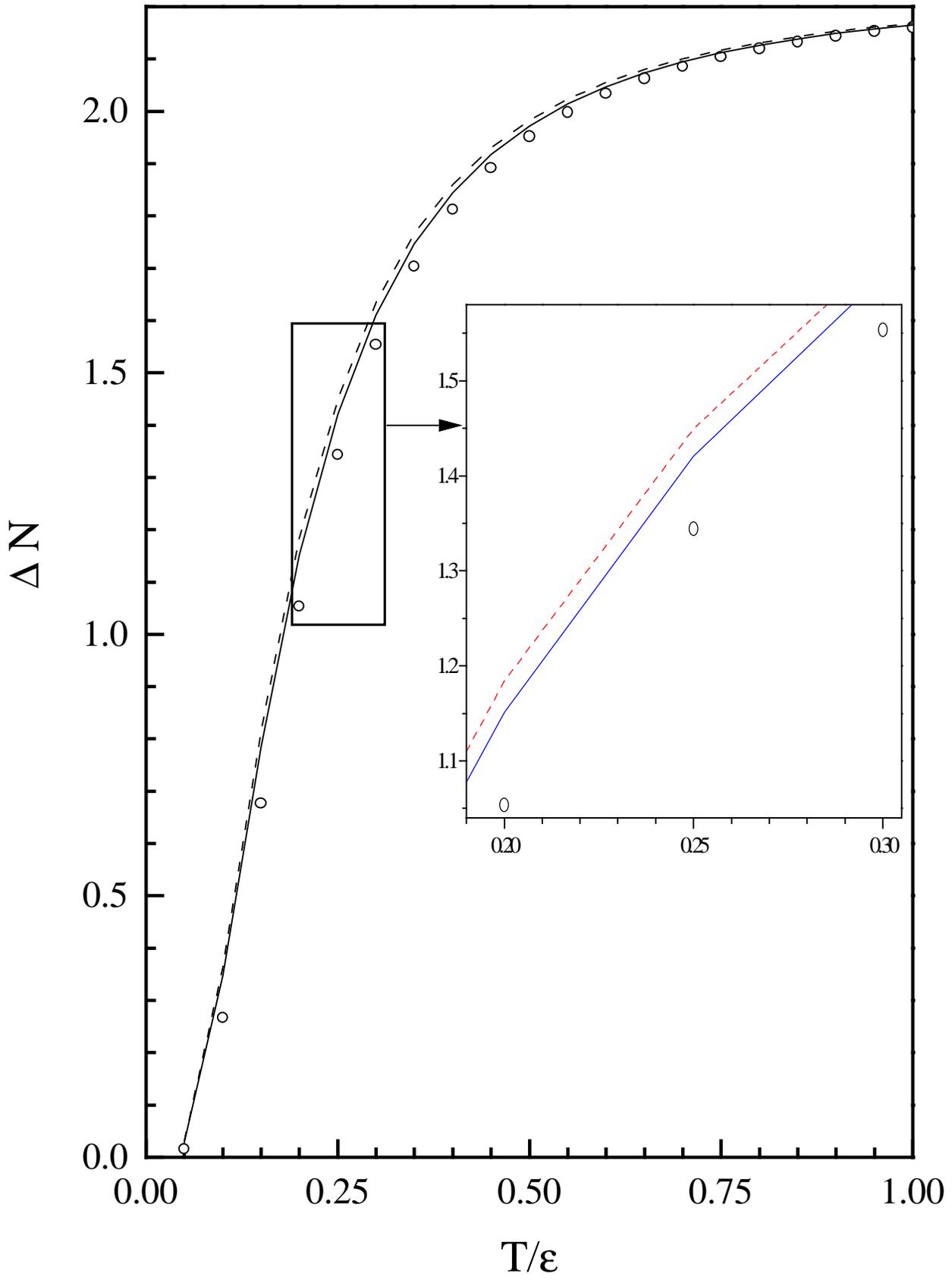
\item[Fig.~5]
The particle number variance $\Delta N$ as a function of temperature
$T$. For notation see Fig.~3.

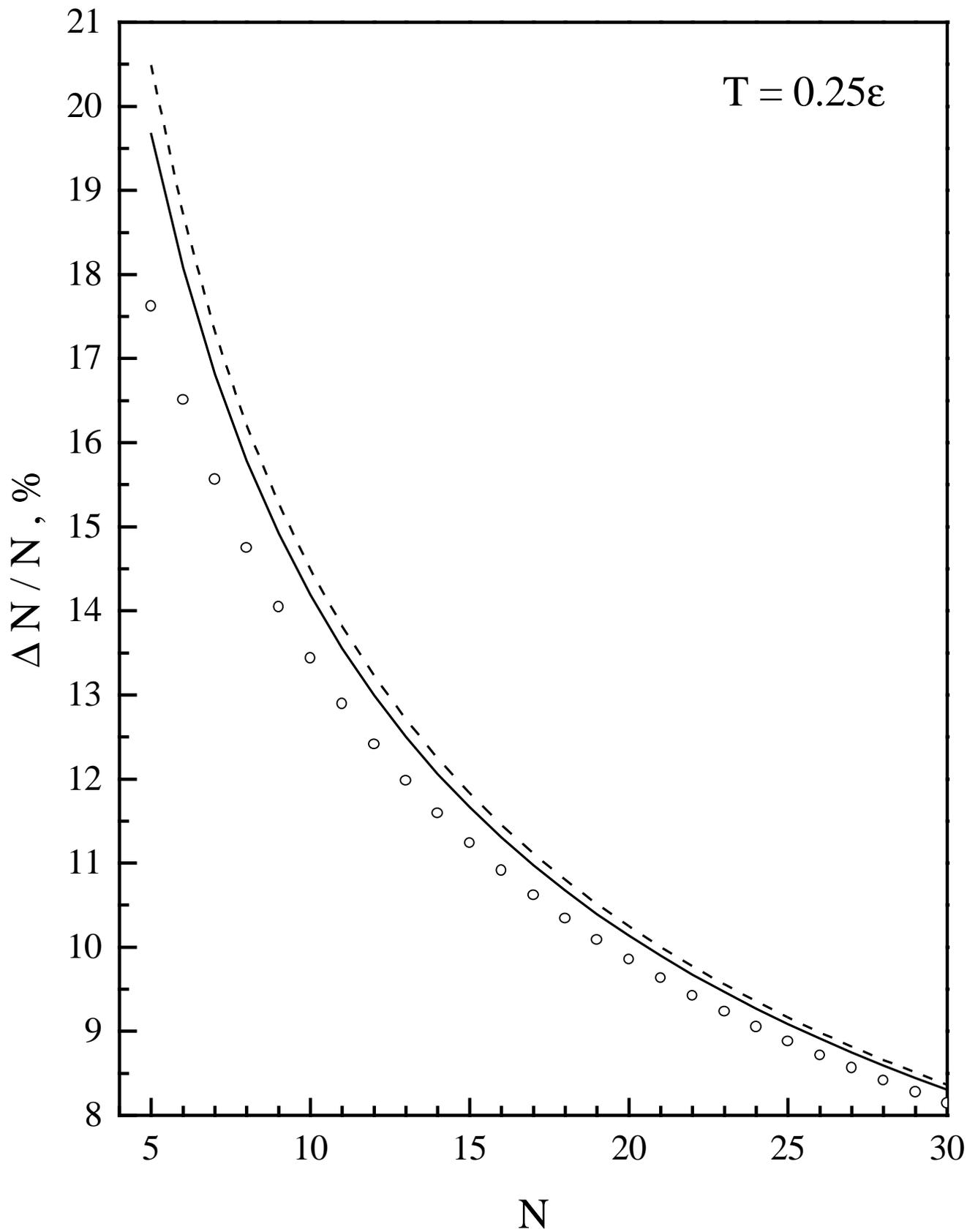
\item[Fig.~6]
The dependence of $\Delta N/N $ on a particle number $N$.
 For notation see Fig.~3.
\end{itemize}

\end{document}